%% file: main.tex
\documentclass[10pt,conference]{IEEEtran}

\newif\ificseanonymous
\IEEEoverridecommandlockouts

\usepackage[T1]{fontenc}
\usepackage{cite}
\usepackage{multirow}
\usepackage{graphicx}
\usepackage{float}
\usepackage{stfloats}
\usepackage{array}
\usepackage{makecell}
\usepackage[most]{tcolorbox}
\usepackage{booktabs}
\usepackage{longtable}
\usepackage{ragged2e}
\usepackage{needspace}
\usepackage{enumitem}
\usepackage{tikz}
\usepackage{xurl}
\usepackage{comment}
\usepackage{balance}
\usepackage[hidelinks]{hyperref}
\usetikzlibrary{positioning, backgrounds, fit, calc}
\DeclareUrlCommand\rr{\urlstyle{tt}}

\AtBeginDocument{
  }

\providecommand{\showeprint}[2][]{\unskip}

\newtcolorbox{takeaway}{
  breakable, enhanced,
  colback=black!3, colframe=black!40,
  boxrule=0.4pt, arc=3pt,
  left=6pt, right=6pt, top=4pt, bottom=4pt,
  before skip=6pt, after skip=6pt,
  title=Summary,
  fonttitle=\bfseries\small,
  titlerule=0pt,
  coltitle=black,
  colbacktitle=black!5,
  attach title to upper={\par\smallskip},
}

\begin{document}
\bstctlcite{IEEEexample:BSTcontrol}

\title{AI Cannot Review What Humans Catch: A First Look at the Reviews of Agent Instruction Files}

\ificseanonymous
  \author{\IEEEauthorblockN{Anonymous Author(s)}
  \IEEEauthorblockA{Anonymous Institution(s)}}
\else
  \author{
    \IEEEauthorblockN{Wenhao Yang}
    \IEEEauthorblockA{
      Peking University\\
      Beijing, China\\
      yangwh@stu.pku.edu.cn}
    \and
    \IEEEauthorblockN{Runzhi He}
    \IEEEauthorblockA{
      Peking University\\
      Beijing, China\\
      rzhe@pku.edu.cn}
    \and
    \IEEEauthorblockN{
      Minghui Zhou%
      \thanks{Corresponding author: Minghui Zhou
      (zhmh@pku.edu.cn).}
    }
    \IEEEauthorblockA{
      Peking University\\
      Beijing, China\\
      zhmh@pku.edu.cn}
  }
\fi

\maketitle

\begin{abstract}
\input{sections/00-abstract.tex}
\end{abstract}

\begin{IEEEkeywords}
coding agents, agent instruction files, code review, large language models, empirical software engineering, open source software
\end{IEEEkeywords}

\input{sections/01-introduction.tex}

\input{sections/02-related-work.tex}

\input{sections/03-methodology.tex}
\input{sections/04-rq1.tex}
\input{sections/05-rq2.tex}
\input{sections/06-rq3.tex}
\input{sections/07-rq4.tex}

\input{sections/09-discussion.tex}

\input{sections/08-threats.tex}
\input{sections/10-conclusion.tex}

\bibliographystyle{IEEEtran}
\balance
\bibliography{reference}
\balance

\end{document}

%% file: sections/00-abstract.tex
Agents increasingly shape software, and instruction files steer them. Files such as \texttt{AGENTS.md}, \texttt{CLAUDE.md}, \texttt{.cursorrules}, and \texttt{SKILL.md} act as executable policy: prose that a machine reads and acts on. Unlike ordinary documentation, a single edit to one of these files can silently redirect how an agent behaves across an entire repository. Natural-language rules have no per-change oracle, so they slip past the compilers, type-checkers, and test suites that guard code. Review is therefore the only quality gate, but no prior work has studied how that review happens.

We analyzed 2{,}870 instruction-file pull requests across 2{,}246 repositories. The gate is largely broken: \textbf{over 70\% of instruction-file changes receive no substantive review}. When an AI reviewer substitutes for a human, it is \emph{structurally blind} to the concerns that most need human judgment. It catches surface structure but misses factual errors, necessity questions, and staleness problems. We contribute the first taxonomy of instruction-file review concerns (16 categories), which exposes this sharp human/AI divergence, and a benchmark of 1{,}962 human-checked concerns across 468 PRs.
We ran three widely-used agents (Codex, OpenCode, and Claude Code) against the benchmark, giving each reviewer the full repository at review time. They recovered \textbf{only 1.3--3.9\% of the concerns humans raise}. Even the strongest current agents cannot substitute for human attention on these files.

%% file: sections/01-introduction.tex
\vspace{-.02in}
\section{Introduction}
\label{sec:intro}

Files such as \texttt{AGENTS.md}, \texttt{CLAUDE.md}, \texttt{.cursorrules}, and \texttt{SKILL.md} tell an agent how to behave inside a repository: which conventions to honor, which commands to run, which paths never to touch. In just a few years they have become a de-facto standard for agent-assisted development: among GitHub repositories that configure an agentic coding tool, 90.6\% ship such a Markdown instruction file~\cite{galster2026configuring}. These files are \textbf{executable policy written in natural language}. They are ordinary prose, carrying none of the formal structure that makes code checkable. A machine reads them as instructions and acts on them. Because the agent acts on that prose literally, a single careless line changes what the agent \emph{does}, not merely how the file reads. That change then propagates into everything the agent does.
Such prose can be wrong in ways that matter: an \texttt{AGENTS.md} rule
that directs the agent to build with a Go release that did not yet exist when the rule was written~\cite{diggerpr2110}, or a checked-in agent-permission file that grants a shell tool whose \texttt{-exec} parameter a maintainer warned could be ``leveraged by a prompt injection''~\cite{keycloakpr47024}.

Unlike the code they govern, agent instruction files escape the automated checks that guard software. Compilers, type-checkers, linters, and test suites encode a checkable definition of correctness and reject faulty changes before review. A natural-language rule has no comparable per-change oracle for whether it is true, necessary, current, or safe~\cite{barr2015oracle,schulhoff2024promptreport}. Prompt- and agent-evaluation frameworks assess only what an instruction set \emph{causes} an agent to do over a fixed task suite~\cite{gloaguen2026evaluatingagents}, not whether a given edit is sound. Review is therefore the only quality gate for these files, yet no one has studied how their changes are reviewed. Understanding that landscape is a first step toward governing and improving these AI practices.

We address this with a census of 2{,}870 instruction-file pull requests across 2{,}246 popular repositories. We find that this gate is largely broken in two ways: \textbf{human review is mostly absent}, and \textbf{where an AI reviewer substitutes for a human, it is structurally blind} to the concerns that most need human judgment. We ask four research questions.

\textbf{RQ1 (Origin): Who authors instruction file changes, and how do they enter the repository?}  Finding: Instruction-file changes increasingly originate from AI and rarely arrive alone. Roughly one-third of the PRs are AI-authored, and 69\% ride along inside a larger PR, dividing the reviewer's attention.

\textbf{RQ2 (Coverage): To what extent and by whom are instruction file changes reviewed?} Finding: 71.9\% of changes receive no substantive review. Where review does occur, humans give rubber-stamp approvals 18.9\% of the time and AI reviews are superficial (review-shaped walkthroughs that render no evaluative judgment) 39.5\% of the time. Even among AI-authored changes, 78\% still pull a human in to make the judgments the AI does not.

\textbf{RQ3 (Divergence): what concerns do human and AI reviewers raise, and how do they diverge?} Finding: Human and AI reviewers do not merely diverge: AI is \emph{structurally blind} to the judgment-dependent concerns humans raise. Humans focus on completeness and factual correctness (24.6\% of their concerns vs 5.4\% of AI's), while AI gravitates to surface structure (19.0\% vs 2.8\%). The concerns only humans raise (unwarranted, stale, and hallucinated content) are almost never caught by AI.

\textbf{RQ4 (Recovery): What share of human concerns can AI reviewers recover?} Finding: Against a benchmark of 1{,}962 verified concerns across 468 PRs, three widely-used agents (Codex, OpenCode, and Claude Code) recover only \textbf{1.3--3.9\%}. Each agent received the repository at review time, so a miss cannot be blamed on missing context. They miss precisely the judgment-dependent categories that humans raise. Because the benchmark captures exactly what AI misses, it is the grounding a future reviewer for these files would need; we sketch such a tool
as future work.

This paper makes the following contributions.
\begin{itemize}[leftmargin=1.2em,itemsep=2pt,topsep=2pt]
  \item The first empirical census of how agent instruction files are reviewed (2{,}870 PRs), showing that review is largely absent and, where present, superficial;
  \item The first taxonomy of instruction-file review concerns (16 categories), which exposes that AI reviewers are structurally blind to the judgment-dependent concerns humans raise; and
  \item A benchmark of 1{,}962 manual validated concerns showing that even the strongest current agents recover only 1.3--3.9\%, evidence that AI cannot substitute for human attention on reviewing agent instruction files.
\end{itemize}

%% file: sections/02-related-work.tex
\section{Background and related work}
\label{sec:related}

\subsection{Agent Instruction Files}
\label{subsec:artifact}

\textbf{Content and adoption.}
Prior work taxonomizes instruction file content across Agent READMEs, \texttt{CLAUDE.md}, Cursor rules, and agent manifests~\cite{chatlatanagulchai2025agentreadmes,chatlatanagulchai2025manifests,mohsenimofidi2025contextengineering,jiang2026cursorrules,santos2025decoding}.
Parallel work studies prompt defects and system-prompt use~\cite{tian2025promptdefects}.
Galster et al.\ report that 90.6\% of agentic-tool-configured repositories ship instruction files~\cite{galster2026configuring}.
Robbes et al.\ estimate 22--29\% coding-agent adoption across 128k projects~\cite{robbes2026agenticmuch}.
Arabat and Sayagh~\cite{arabat2026instructionsascode} treat instruction files as software artifacts that should be reviewed, measuring merge rate, code churn, and the number of review comments each PR draws.

\textbf{Quality, security, and governance.}
Instruction file quality matters because defects alter what a non-deterministic agent does.
Context length and prompt formatting influence model behavior~\cite{sclar2024sensitivity,liu2024lostmiddle}.
Gloaguen et al.~\cite{gloaguen2026evaluatingagents} find that repository-level context files do not significantly improve task-success rates on SWE-bench or their own CTXBENCH benchmark.
They also raise inference cost by over 20\%, and LLM-generated files sometimes reduce success outright.
Instructions also pose prompt-injection risks in agentic editors~\cite{liu2025aishelljack}.
Security-smell work on infrastructure configuration provides the closest analogy for systematically reviewing privileged instructions~\cite{rahman2019sevensins}.
Researchers studying prompt defects and governance frame these files as policy-bearing artifacts with specification, structure, and community-governance risks~\cite{tian2025promptdefects,yang2026genaigovernance}.

\subsection{Reviewing Software Artifacts}
\label{subsec:reviewing}

\textbf{Code review.}
Classic studies sort review comments into functional defects and maintainability issues such as readability, naming, and structure.
They find that maintainability concerns dominate, accounting for roughly 75\% of comments~\cite{mantyla2009defects,beller2014modern}.
Turzo and Bosu~\cite{turzo2024reviewcomments} built a taxonomy of review comments.
Other work studies why reviewers comment, which comments lead authors to make changes, and how review transfers knowledge between developers~\cite{li2017reviewcomments,bacchelli2013modernreview,bosu2015useful}.

\textbf{Documentation and configuration review.}
Previous work on documentation reviews studies issue types, outdated references, installation guidance, documentation-review comments, and README linting~\cite{aghajani2019documentation,tan2024outdatedreferences,gao2025installation,rao2022commentsoncomments,mynampaty2026lintme}.
Work on configuration distinguishes two failure modes we borrow.
\emph{Misconfiguration} means wrong parameter values that break the system; Xu et al.\ argue this is a design problem rather than a user problem~\cite{xu2013misconfigurations}.
\emph{Configuration smells} are maintainability and security anti-patterns in configuration scripts that a review can catch even when the values themselves are valid~\cite{zhang2021configuration,sharma2016configsmell,rahman2021securitysmells}.

\textbf{Substantive review.}
Another strand of studies focuses on the rubber-stamp effect of reviews: changes are approved without scrutiny.
Look good to me (LGTM) review smell studies report 64.7\% comment-free rate and 18\% LGTM rate in pull requests~\cite{dogan2022reviewsmells,gon2024lgtm}.
Similar work quantifies a shallow review rate of 14\%~\cite{chouchen2021shallowreview}.
Bot participation and differences between human and LLM comment categories provide additional baselines~\cite{wessel2022qualitygatekeepers,goldman2025resolvedcomments}.

\textbf{AI reviewers and benchmarks.}
Both industry and academia have invested in automated review generation since the pre-LLM era~\cite{tufano2021automatingreview,lu2023llamareviewer,hong2025ragreviewer,sun2025bitsaicr,github2026copilotreviews}.
GitHub reports 60M Copilot reviews in production~\cite{github2026copilotreviews}, a vendor-reported scale signal.
Evaluations of automated code review have moved from text-similarity metrics toward recall against human review comments~\cite{zverianskii2026codereviewbench,pereira2026crbench,naik2025crscore,lu2025deepcrceval,voorhees2003trecqa,pradeep2025greatnuggetrecall}.

\subsection{The Knowledge Gap}
\label{subsec:positioning}

Instruction files now steer much of what software agents do, and they are widely adopted and consequential.
Yet they lack strong validation; review is the main quality gate for each change.
Researchers have studied code review, documentation review, and configuration review extensively, but not changes to instruction files.
These files have no per-change oracle, and no prior work provides the comment-level census, taxonomy, or AI benchmark we build here.

%% file: sections/03-methodology.tex
\section{Methodology}
\label{sec:method}

\begin{figure}[t]
  \centering
  \includegraphics[width=\columnwidth]{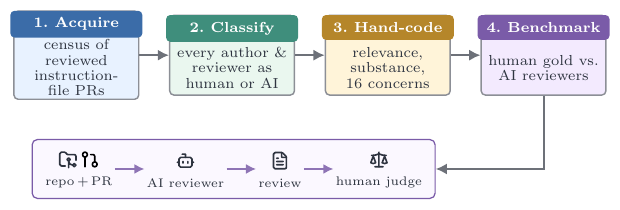}
  \caption{Overview of the study pipeline.}
  \label{fig:overview}
\end{figure}

Instruction files have no per-change oracle, and no existing dataset separates instruction-file review from ordinary code review. A large-scale, hand-validated census of real pull requests is therefore the only practical way to see how these files are reviewed. Surveys would burden already-overloaded maintainers and still miss scale; case studies risk project-specific bias and cannot capture the broader landscape.
We mine the whole of GitHub and reconstruct the review lineage for thousands of agent-instruction-file PRs.
The study has four stages (Figure~\ref{fig:overview}):
\textbf{(1) acquire} a census of reviewed instruction file PRs; \textbf{(2) classify} every actor as human or AI; \textbf{(3) hand-code} each relevant review act; and \textbf{(4) validate and benchmark} the human concerns against product-native AI reviewers.

\subsection{Data acquisition \& the corpus}
\label{subsec:acquire}

We assemble the corpus in three stages.
\textbf{(1) Candidate discovery.} We first query the April~2026 snapshot of World of Code~\cite{ma2021worldcode} for commits whose changed paths match our instruction-file patterns ($\approx$460k commits).
We then map these commits to GitHub pull requests using GHArchive event data~\cite{gharchive2026} and retain only candidates from repositories with more than 100~stars, yielding 3,765 candidate PRs.
\textbf{(2) PR retrieval.} Archival datasets contain noise and hollowing~\cite{kalliamvakou2014github}, so we cross-validate each candidate against the GitHub API. We keep only pull requests whose full metadata, changed files, reviews, and comments can be recovered.
\textbf{(3) Static snapshot.} For each verified PR, we compile a static snapshot: PR metadata, changed files, review and discussion comments, review records, and the base and head versions of each changed instruction file.
A PR enters the corpus only if it (i) modifies at least one instruction file, (ii) contains at least one non-author review comment in the recovered GitHub data, and (iii) can be consistently reconstructed at the reviewed revisions.
The corpus contains 2,870 verified instruction-file PRs from 2,246 repositories. These PRs yield 5,475 instruction-file instances, where each instance is one changed instruction file within one PR.

Table~\ref{tab:repo-demographics} summarizes the 2,246 corpus repositories. The four metrics are heavily skewed, so we report mean and median and visualize each distribution.

\begin{table}[t]
\centering
\caption{Demographics of the corpus repositories.}
\label{tab:repo-demographics}

\begin{tabular}{@{}l rr c@{}}
\toprule
\textbf{Statistics} & \textbf{Mean} & \textbf{Median} & \textbf{Distribution} \\
\midrule
\# of Stars & 8,565.83 & 1,426.00 & \raisebox{-0.3\height}{\includegraphics[width=1.6cm,height=0.45cm]{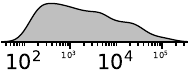}} \\
\# of Commits & 8,738.32 & 2,198.50 & \raisebox{-0.3\height}{\includegraphics[width=1.6cm,height=0.45cm]{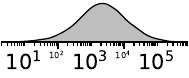}} \\
\# of Contributors & 246.27 & 56.00 & \raisebox{-0.3\height}{\includegraphics[width=1.6cm,height=0.45cm]{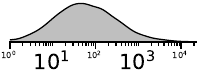}} \\
Age (years) & 5.99 & 5.26 & \raisebox{-0.3\height}{\includegraphics[width=1.6cm,height=0.45cm]{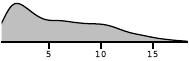}} \\
\midrule
Language & \multicolumn{3}{l}{TS (27.4\%), Py (18.7\%), Go (8.4\%), Other (45.5\%)} \\
\addlinespace[4pt]
Owner type & \multicolumn{3}{l}{Org (1,706), User (540)} \\
\bottomrule
\end{tabular}
\end{table}

\subsection{Actor classification --- author \& reviewer}
\label{subsec:actor}

Before measuring anything else, we label every actor in the corpus.

\textbf{Author origin (RQ1).} For each change, we infer its author origin from structured signals in the PR artifact and from textual evidence in the PR description and discussion. We distinguish four mutually exclusive origins: (i) \emph{pure-human}, (ii) \emph{AI-generated-declared}, (iii) \emph{AI-agent/bot}, and (iv) \emph{self-admitted-AI-assisted}.
AI-generated-declared and AI-agent/bot changes carry unambiguous machine-emitted markers in the PR metadata, specifically fixed co-author lines or generated-by trailers from tools such as Claude Code, Copilot, or Codex. Self-admitted-AI-assisted changes come from free-text self-reports in the PR description or comments (e.g., ``vibe-coded'' or ``with help of \ldots''). These self-reports are noisier but still explicit. Changes that show none of the above signals are labeled pure-human.
To validate the labeling procedure, we manually audited a stratified random sample of 150 PRs. We found a missed-AI-attribution rate of 2.0\% (3 errors). Because our method relies on explicit declarations and may miss unreported AI involvement, the AI-authored share we report is a conservative lower bound.

\textbf{Reviewer role (RQ2--RQ4).} We automatically classify each reviewer account as Human, AI, or Tool based on signals drawn from public account metadata. We exclude reviewer accounts whose self-reported status says they skipped or failed the review. We validate the classifier against a manually curated gold standard of 300 stratified reviewer accounts, independently coded by two annotators (Cohen's $\kappa = 0.983$). The classifier achieves 95.3\% accuracy under account-level weighting, where each unique reviewer account contributes once. It achieves 98.5\% accuracy under PR-instance-level weighting, where each appearance of a reviewer in a PR is counted separately.

\subsection{Human coding}
\label{subsec:coding}

\begin{figure}[!t]
  \centering
  \includegraphics[width=\columnwidth]{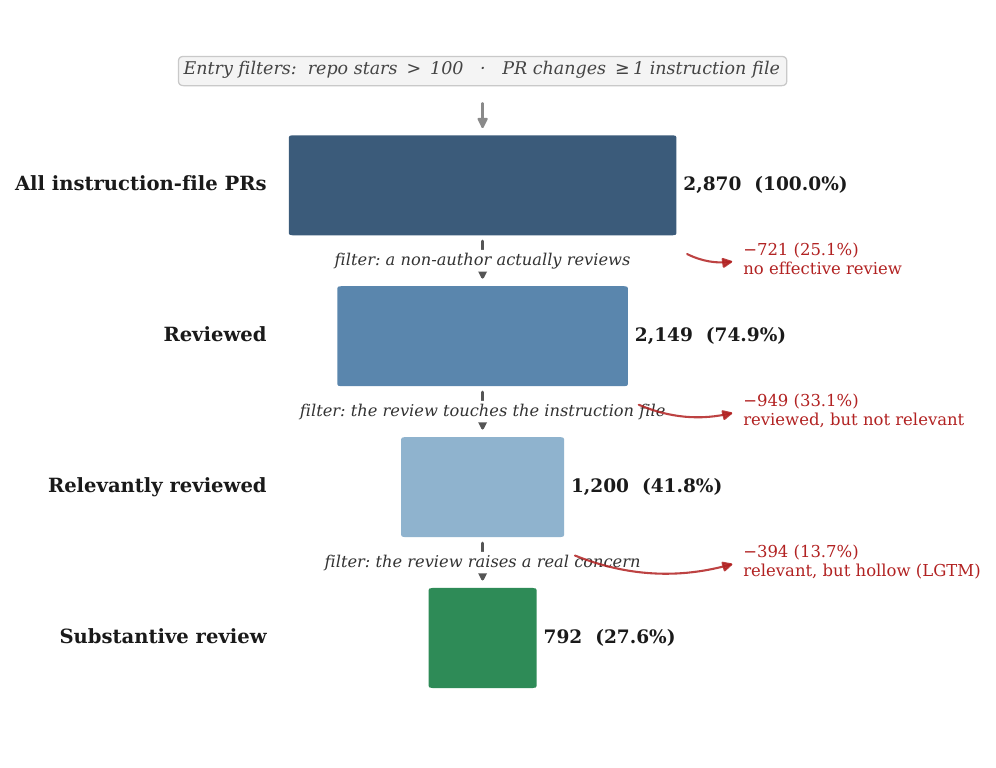}
  \caption{The review funnel over 2,870 instruction-file PRs.}
  \label{fig:funnel}
\end{figure}

\textbf{Relevance (automated + human).} A comment is relevant if it addresses a changed instruction artifact. We first apply two deterministic rules. Together they cover 91.8\% of all non-author review comments (21,710 of 23,651). Rule~(1): an inline review comment left \emph{on the changed instruction file}, such as a diff-anchored suggestion on a specific line. Rule~(2): a bare approval in a \emph{pure-instruction} PR that changes only instruction files, such as an ``LGTM'' with no inline comments at all. We also exclude empty bodies and status-only bot noise.
The remaining 8.2\% of comments are hand-coded and adjudicated. Inter-coder agreement was moderate ($\kappa = 0.630$), reflecting the inherent ambiguity of relevance at the boundary between instruction-file and non-instruction-file discussion. Aggregating to the PR level, the funnel over all 2,870 PRs is: 721 with no effective non-author review (no human or AI reviewer left a comment after we exclude bot skip/fail messages), 949 reviewed but not relevant, and 1,200 relevantly reviewed (Figure~\ref{fig:funnel}).

\textbf{Substantive \& concern coding (inductive thematic analysis).}
\label{subsec:substantive}
From the 1,200 relevantly-reviewed PRs, a single coding scheme yields both the concern taxonomy and a substantive/non-substantive label. A comment is substantive iff the coding attaches at least one concern, i.e., a genuine evaluative point about the change's content, placement, naming, or governance. Two authors inductively open-coded approximately 300 PRs, iterated to a stable codebook, and then double-coded 99.3\% of the 1,200 PRs. The substantive-vs-non-substantive \emph{tier-gate}, which marks whether a comment carries any concern, reached $\kappa = .918$. Per-concern code assignments agree at a macro-average $\kappa = .72$; most codes fall between .71 and .89, and a few judgment-heavy codes dip lower. Disagreements were adjudicated to gold, yielding 1,987 substantive comments across the 1,200 PRs.

\textbf{Identity-blind coding.}
\label{subsec:blind}
The substantive/concern pass is identity-blind. Coders never see reviewer login, class, or mix; they see only the instruction file diff and comment texts. Reviewer identity is joined back only after labels are frozen.

\subsection{Validation \& evaluation}
\label{subsec:validate}

\textbf{Validate comments $\rightarrow$ the gold.} The 1,987 substantive comments feed this step. Two coders validate each concern (KEEP/DROP/UNSURE) by reading the PR, inspecting the current repository, and searching the web. They judge each concern against three tiers of evidence, from strongest to weakest: \emph{outcome} (did the author later change the flagged content?), \emph{discourse} (a ``good catch'' or ``fixed'' reply vs.\ a correct author rebuttal), and \emph{factual truth} (is the claim itself correct, checked against the code and docs?). Double-coding and adjudication trim 25 confirmed false positives (1.26\% of the 1,987 substantive comments). The result is a gold set of 1,962 concerns across 468 PRs.

\textbf{Benchmark (reviewer execution \& coverage judging).}
\label{subsec:benchmark}
We reconstruct each PR at its review-time state and invoke each product's native review on the full repository (base and head) with the archived PR title and body.
The three subjects are Codex \texttt{review} (GPT-5.5 high), OpenCode \texttt{review} (DeepSeek-v4-pro max), and Claude Code \texttt{/review} (Claude Sonnet 4.6 high). We run each system on 446 of the 468 gold PRs that could be reconstructed at review time; the remaining 12 PRs were unreconstructable.
We disabled web tools and trimmed git history to prevent leakage.

We measure coverage as strict nugget-recall against the gold: each gold concern is a nugget, and we report what fraction the AI review surfaces. \textbf{Two human judges} independently judge each AI review. A \texttt{covered} verdict requires the judge to supply a contiguous verbatim quote from the AI review that raises the \emph{same concern in the same direction} as the gold concern. A match counts even when the AI's wording or proposed fix differs from the gold; praise or neutral description does not count.

%% file: sections/04-rq1.tex
\section{RQ1: Origin}
\label{sec:rq1}

Before examining how instruction-file changes are reviewed, we characterize the changes themselves: who writes them, whether they travel alone or with other work, and whether they create a file or edit an existing one. These properties shape the task that later review faces. The full census covers 2,870 PRs across 2,246 repositories and contains 5,475 instruction-file instances. The unit of analysis throughout is the PR.

\textbf{Authorship.}\label{subsec:rq1-authorship} We classify each PR by the origin of its instruction-file changes into four mutually exclusive author-origin classes, following the detection method of Section~\ref{sec:method}. Roughly one third are explicitly generated or assisted by AI. Table~\ref{tab:rq1-authorship} shows the breakdown: \textbf{963} of 2,870 PRs fall into one of the three AI-associated classes, while the rest are pure human. These counts capture only cases with \emph{explicit} AI evidence (a self-declaration, a machine-emitted trailer, or a bot author account). Undisclosed AI use is not counted, so the true AI share is likely higher.

\begin{table}[t]
  \centering
  \caption{Author-origin classes.}
  \label{tab:rq1-authorship}
  \small
  \begin{tabular}{lrr}
    \toprule
    Author origin & PRs & of 2,870 \\
    \midrule
    Pure human & 1,907 & 66.4\% \\
    AI-generated (declared) & 403 & 14.0\% \\
    Human, self-reported AI use & 192 & 6.7\% \\
    AI agent / bot & 368 & 12.8\% \\
    \bottomrule
  \end{tabular}
\end{table}

Authorship tells us who writes the changes. We now ask how they enter the repository: alone, or bundled with other work that competes for reviewer attention.

\textbf{Ride-along.}\label{subsec:rq1-ridealong} Instruction changes rarely arrive alone. We call a PR a \emph{rider} if it changes at least one non-instruction file and \emph{standalone} otherwise. \textbf{1,979 (69.0\%)} are riders. Riders carry a median of 4 other files. The distribution has a heavy right tail: a handful of mega-PRs sweep up thousands of vendored, generated, or lockfile entries, and the largest touches 1.34M lines. Among riders, the instruction edit most often travels with code (68.7\%), then docs (58.2\%) and config (51.0\%).

\begin{table}[t]
  \centering
  \caption{Co-changed file types in rider PRs.}
  \label{tab:rq1-ridealong}
  \small
  \begin{tabular}{lrr@{\hskip 3em}lrr}
    \toprule
    type & PRs & share & type & PRs & share \\
    \midrule
    code   & 1,360 & 68.7\% & build/deps  & 778 & 39.3\% \\
    docs   & 1,152 & 58.2\% & test        & 776 & 39.2\% \\
    config & 1,010 & 51.0\% & CI/workflow & 503 & 25.4\% \\
    other  &   802 & 40.5\% & lockfile    & 499 & 25.2\% \\
    \bottomrule
  \end{tabular}
\end{table}

\textbf{Change type.}\label{subsec:rq1-changetype} The dominant change is adding a new instruction file, not editing an existing one. By instances, 72.4\% of the 5,475 instruction-file occurrences are added as new. By PRs, \textbf{1,925 (67.1\%)} are \emph{pure-add}: they only add instruction content, with no deletions to existing instructions. Most instruction files enter the repository as a brand-new addition. The reviewer has no prior version to diff against and must evaluate the content on its own terms.

\begin{takeaway}
Roughly a third of instruction-file changes are AI-authored (963/2,870), 69\% ride along inside larger code PRs rather than arriving alone, and 67\% add a brand-new file with no prior version to diff against, three conditions that together leave the reviewer no baseline to judge the change against.
\end{takeaway}

%% file: sections/05-rq2.tex
\section{RQ2: Coverage}
\label{sec:rq2}

RQ2 asks three questions: who reviews instruction-file changes, whether any review reaches the change, and whether that review offers a real evaluative judgment. We call the last failure \emph{review hollowing}: review is present but never engages with or evaluates the instruction-file change.

\subsection{Who reviews the changes}
\label{subsec:rq2-who}

Table~\ref{tab:rq2-mix} shows the reviewer mix for the 2,870 instruction-file PRs (classification in Section~\ref{sec:method}).

\begin{table}[t]
  \centering
  \caption{Reviewer mix across the two review gates.}
  \label{tab:rq2-mix}
  \small
  \begin{tabular}{lrrr}
    \toprule
    Reviewer mix & PRs & Relevant ($n/N$) & Substantive ($n/N$) \\
    \midrule
    silent      & 717 & 0/717        & ---            \\
    human-only  & 882 & 411/882     & 242/411   \\
    ai-only     & 484 & 337/484     & 194/337   \\
    human+ai    & 530 & 446/530     & 353/446   \\
    tool-only   & 257 & 6/257       & 3/6       \\
    \midrule
    Total       & 2,870 & 1,200/2,870 & 792/1,200 \\
    \bottomrule
  \end{tabular}
\end{table}

\textbf{AI both writes and reviews.} AI authors about one-third of these changes (RQ1) and reviews 1,014 of the 2,870 PRs (35.3\%; the ai-only and human+ai splits appear in Table~\ref{tab:rq2-mix}). These two roles overlap. Among the 2,675 PRs jointly covered by the authorship and reviewer analyses (93\% of the census), 417 (15.6\%) are both AI-written and AI-reviewed. In 148 (5.5\%), an AI reviewer is the only reviewer present, with no human in the loop; we examine that configuration in RQ4.

\subsection{Relevance: did any review touch the instruction change?}
\label{subsec:rq2-relevance}

The first gate asks whether any non-author comment touched the instruction-file change. The relevance screen splits the 2,870 PRs into three groups: 721 with no effective non-author review, 949 reviewed-but-not-relevant, and 1,200 relevantly reviewed. Overall, 58.2\% of instruction-file PRs receive no relevant comment at all, and only 1,200 (41.8\%) are relevantly reviewed. AI reviewers reach the instruction file more often than humans do, and human+ai pairs reach it most often (Table~\ref{tab:rq2-mix}). Silent and tool-only PRs draw essentially no relevant comments.

\textbf{Reviewed-but-not-relevant is a failure mode prior work cannot see.} The 949 PRs (33.1\%) draw review on the PR's code but not on its instruction file. Prior shallow-review work would count these simply as ``reviewed'', because it tracks whether a PR received any review, not whether that review touched the instruction file. This matches RQ1: many instruction edits ride along on code-centric PRs, so reviewers focus on the code and ignore the instruction change.

\subsection{Substantive: did the review raise a real concern?}
\label{subsec:rq2-substantive}

We now examine individual comments. Reviewers left 4,789 comments on the 1,200 relevant PRs. We classify each comment as \emph{substantive} if it raises at least one evaluative concern, or \emph{hollow} if it touches the file without rendering an evaluative judgment. Overall, 66.9\% are substantive and 33.1\% are hollow.

\textbf{The human/AI divergence.} Humans are substantive \textbf{73.9\%} of the time; AI reviewers are substantive only \textbf{57.4\%}. The two also hollow out in different ways.

\textbf{The two hollow failure modes are qualitatively distinct.} Human hollow review is usually a rubber-stamp approval (18.9\% of human comments), an ``LGTM''-style sign-off. AI hollow review is almost always a walkthrough / PR-summary (39.5\%). It looks like review, a tidy file-by-file recap, but renders no evaluative judgment.

\textbf{The review funnel.} Combining both gates at the PR level gives the funnel in Figure~\ref{fig:funnel}. Across all 2,870 instruction-file PRs, 71.9\% receive no substantive review of the instruction change; only 792/2,870 (27.6\%) do.

\begin{takeaway}
Only 41.8\% (1,200/2,870) of instruction-file changes draw a comment that even touches the instruction file, and just 27.6\% (792) draw a substantive one; over 70\% clear the gate with no real evaluation. Most of the shortfall is silence or code-only review; even among reviews that do reach the file, a third of the comments are hollow ``LGTM'' or walkthrough sign-offs that render no judgment.
\end{takeaway}

%% file: sections/06-rq3.tex
\section{RQ3: Divergence}
\label{sec:rq3}

We hand-code the substantive review comments and identify \textbf{16 concerns} (Table~\ref{tab:rq3-taxonomy}), the quality issues a reviewer raises about an instruction-file change. Human and AI reviewers distribute across these concerns very differently. Instruction files have \emph{no per-change oracle}: no automated ground truth that says whether a given directive is true, necessary, current, or safe. Because nothing forces reviewers to converge on a single ``should catch'' list, human and AI reviewers act as non-redundant checkers with systematically different attention patterns. Humans flag what only a project insider can judge (whether a directive is true, complete, warranted, or out of date), while AI reviewers disproportionately stay at the surface of formatting and embedded-code syntax. That contrast is the paper's central empirical claim.

\begin{table*}[!tp]
\centering
\caption{The instruction-file review-concern taxonomy.}
\label{tab:rq3-taxonomy}
\footnotesize
\setlength{\tabcolsep}{4pt}
\begin{tabular}{@{}>{\raggedright\arraybackslash}m{2.5cm}m{4.5cm}m{7.2cm}r r r@{}}
\toprule
Concern & What the reviewer questions & Example & $n$ & \makecell[r]{\% of\\human} & \makecell[r]{\% of\\AI} \\
\midrule
\multicolumn{6}{@{}l}{\small\bfseries Presentation} \\
\addlinespace[3pt]
Structure / formatting & Spelling, grammar, Markdown, or layout. & AI at \rr{primer/react#6071}: ``there's a typo in `multpile'; it should be `multiple'.'' & 489 & 2.8 & \textbf{19.0} \\
\midrule
\multicolumn{6}{@{}l}{\small\bfseries Correctness} \\
\addlinespace[3pt]
Executability & A command, path, link, or reference will not resolve or run as written. & AI at \rr{tambo-ai/tambo#1116}: ``Read Docs/AGENTS.md'' uses the wrong casing and is not a valid link. & 311 & 5.1 & \textbf{8.2} \\
Consistency & Conflicts with another rule, a named in-project source, or its own example. & AI at \rr{k3nsei/ng-in-viewport#1556}: ``update the Node.js requirement\ldots\ to match \texttt{package.json} engines.'' & 359 & 4.5 & \textbf{11.6} \\
Factual accuracy & A stated fact about the project, toolchain, versions, or workflow is wrong as written. & Human at \rr{openfoodfacts/openfoodfacts-server#12271}: ``the \texttt{checks} target runs a full frontend build and frontend lint, too''---correcting what the file claims the command does. & 262 & \textbf{7.8} & 2.5 \\
Stale / outdated & Once valid, the content has aged out as versions, APIs, or practices changed. & Human at \rr{decidim/decidim#15187}: ``this shouldn't happen anymore after the rbenv introduction.'' & 78 & \textbf{2.8} & 0.2 \\
Hallucinated / fabricated & Refers to a file, command, or practice that never existed here. & Human at \rr{PrivateBin/PrivateBin#1698}: ``the repo doesn't contain a public directory---I assume this got picked up from our setup hardening instructions?'' & 46 & \textbf{1.5} & 0.3 \\
\midrule
\multicolumn{6}{@{}l}{\small\bfseries Coverage} \\
\addlinespace[3pt]
Conciseness & Valid content that is duplicated or verbose. & AI at \rr{kornia/kornia#3620}: the section ``duplicates the development setup and task list already documented in CONTRIBUTING.md.'' & 103 & \textbf{2.5} & 1.8 \\
Specificity / actionability & Present but too vague or underspecified for an agent to act on. & Human at \rr{dotnet/diagnostics#5609}: ``is `Native component issue' a concept you recognize? \ldots\ could you give an example?'' & 201 & \textbf{4.6} & 3.6 \\
Completeness & A rule, command, prerequisite, or example the agent needs is missing. & Human at \rr{OWASP/mastg#3447}: ``list prerequisites needed to execute or evaluate the test\ldots\ existing files are in \texttt{prerequisites/}.'' & 775 & \textbf{24.6} & 5.4 \\
\midrule
\multicolumn{6}{@{}l}{\small\bfseries Efficacy} \\
\addlinespace[3pt]
Discoverability / loading & Whether the file or rule will actually be found, loaded, or scoped by the agent tooling. & Human at \rr{Pipelex/pipelex#105}: ``please make this an on-demand rule, not \texttt{alwaysApply}.'' & 140 & \textbf{3.2} & 2.5 \\
Wrong mechanism & The behavior should be a hook, linter, or CI check, not natural-language prose. & Human at \rr{langchain-ai/langchain#32075}: ``let's avoid using copilot as a replacement for a linter\ldots\ the linter is already doing this.'' & 23 & \textbf{0.8} & 0.1 \\
Security & Secrets, over-broad permissions, prompt injection, or other unsafe agent behavior. & Human at \rr{langfuse/langfuse-python#1268}: ``we are allowing all \texttt{cat} commands\ldots\ claude might decide to read \texttt{.env}\ldots\ and send to LLM.'' & 79 & 1.6 & \textbf{1.8} \\
\midrule
\multicolumn{6}{@{}l}{\small\bfseries Warrant} \\
\addlinespace[3pt]
Scope / placement & Fine content in the wrong file, directory, or layer. & Human at \rr{buildbuddy-io/buildbuddy#9822}: put general guidance ``in CONTRIBUTING.md instead of adding a bunch of new files specific to only some IDEs.'' & 111 & \textbf{3.6} & 0.6 \\
Audience fit & May be true, but aimed at the wrong reader---a human maintainer, not the agent. & Human at \rr{dashpay/dash#6741}: ``it's not for Claude, it's for the maintainer.'' & 39 & \textbf{1.2} & 0.3 \\
Triviality / low-value & Has no actionable value on its own---obvious or empty. & Human at \rr{jitsi/jitsi-meet#16459}: ``I don't think this does anything useful, it just lists dependencies.'' & 159 & \textbf{5.6} & 0.4 \\
Overreach / unwarranted & Whether the rule or file should exist at all, even if accurate and well formed. & Human at \rr{quarkusio/quarkus#53038}: ``are we \emph{certain} we want AIs to be using \texttt{onlyIf}?'' & 109 & \textbf{3.9} & 0.2 \\
\bottomrule
\end{tabular}
\end{table*}

\subsection{The concern taxonomy}
\label{subsec:rq3-taxonomy}

We code each substantive comment into 16 concerns. Table~\ref{tab:rq3-taxonomy} groups them by the kind of quality questioned and orders them from surface-level, machine-checkable concerns to context-dependent, judgment-heavy ones. The grouping is expository, not a coding axis; a comment may carry up to two concerns. Most of the taxonomy is shared with code and documentation review. Five concerns have no equivalent in prior code-, documentation-, or prompt-review taxonomies: warrant (\emph{Triviality}, \emph{Overreach}), agent efficacy (\emph{Audience fit}, \emph{Discoverability / loading}), and a new AI-authorship subcategory of correctness (\emph{Hallucinated}). These five are the taxonomy's distinctive contribution. The percentage columns show each concern's share \emph{within that reviewer type's own comments} (2{,}687 human / 2{,}042 AI).

\subsection{The human/AI divergence}
\label{subsec:rq3-asymmetry}

The distribution reveals the divergence immediately. The three largest concerns are Completeness ($n{=}775$), Structure ($n{=}489$), and Consistency ($n{=}359$); together they account for the bulk of substantive comments. The long tail contains the judgment and AI-risk concerns: Triviality, Overreach, Stale, Hallucinated, Audience fit, and Wrong-mechanism. These appear rarely, but almost entirely in human comments. Their rarity does not make them unimportant; it means no reviewer or tool is currently raising them.

\begin{figure}[!t]
  \centering
  \includegraphics[width=\columnwidth]{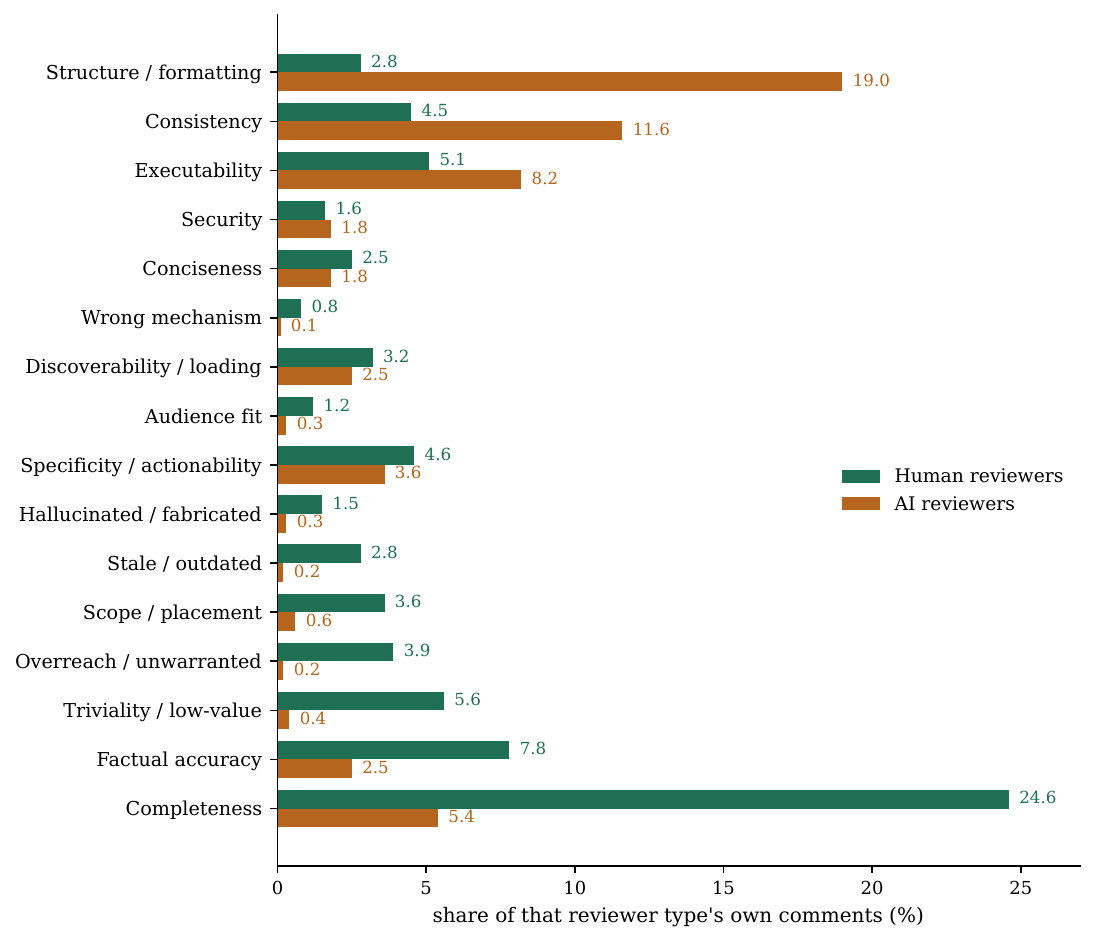}
  \caption{Each concern's share of human vs.\ AI substantive comments.}
  \label{fig:asymmetry}
\end{figure}

Human comments are substantive \textbf{73.9\%} of the time and AI comments \textbf{57.4\%} (the RQ2 split). But \emph{which} concerns each side raises diverges sharply (Table~\ref{tab:rq3-taxonomy} and Figure~\ref{fig:asymmetry}). The concerns fall into three clusters:

AI leads on the machine-checkable surface. Structure / formatting (19.0\% of AI's comments vs.\ 2.8\% of humans'), Consistency with a declared spec (11.6 vs.\ 4.5), and Executability / broken references (8.2 vs.\ 5.1) can all be pattern-matched or run to verify.

Humans lead on the judgments that lack an oracle. They catch what is missing (Completeness, 24.6 vs.\ 5.4), whether it is true (Factual accuracy, 7.8 vs.\ 2.5), whether it adds value (Triviality), whether it oversteps (Overreach), whether it has gone stale (Stale), whether the AI invented it (Hallucinated), and whether the placement, audience, or mechanism is right. These judgments require knowing the project, the intent, and the domain; no automatic checker can make them.

Security, Discoverability, Specificity, and Conciseness form a shared band. Both sides raise boundaries, loading, clarity, and cost at similar rates.

\textbf{The human is the backstop.} The gap is concrete, not hypothetical: when humans and AI review the same file, humans supply the judgments AI misses. Among the 446 PRs reviewed by both, 139 contained an AI-authored instruction file. In \textbf{78\%} of those (109/139) the AI review missed at least one substantive concern that a human then caught. Humans also overrode the AI outright in 13 PRs, making the Overreach and Triviality judgments that AI does not make. On these files AI review is advisory, not authoritative.

Attention is not capability. The patterns above describe what AI \emph{does} raise in the wild, not what a \emph{dedicated} AI \emph{could} raise. RQ4 turns that distinction into a capability test.

\begin{takeaway}
Human and AI reviewers check different things, not the same things at different rates. AI concentrates on surface structure: formatting (19.0\% of its comments vs.\ 2.8\% of humans'), spec consistency, and broken references, while humans dominate the judgments that lack an oracle: what is missing (Completeness, 24.6\% vs.\ 5.4\%), untrue, stale, or unwarranted. Where both review the same AI-authored file, humans catch a concern the AI missed 78\% of the time (109/139); AI judges well-formedness, but whether a rule should exist is invisible to it.
\end{takeaway}

%% file: sections/07-rq4.tex
\section{RQ4: Recovery}\label{sec:rq4}

RQ3's observational distribution cannot settle what AI reviewers \emph{can} raise. We therefore test whether a dedicated AI reviewer, given the same change and full repository, reproduces a human concern (benchmark mechanism in Section~\ref{sec:method}).

\subsection{How much do they recover?}\label{subsec:rq4-part2}

We score the three reviewers against the 1,912 expert-validated human gold concerns (446 PRs); two human judges scored coverage (Section~\ref{sec:method}).

OpenCode recovers 3.9\% (75/1,912), Claude 3.2\% (62/1,912), and Codex 1.3\% (25/1,912); every reviewer misses \textbf{$\approx$96--99\%}. Coverage is scored strictly---the same concern in the same direction, backed by a verbatim quote (nugget recall, Section~\ref{sec:method})---and the ranking is stable under a stricter independent re-judging (3.5\%, 2.9\%, 1.2\%).

Table~\ref{tab:rq4-percat} gives per-category coverage of the gold, grouped from surface-level, machine-checkable concerns (top) to context-dependent, judgment-heavy ones (bottom).

\begin{table}[t]
  \centering
  \caption{Per-category coverage of the 1,912-concern human gold.}
  \label{tab:rq4-percat}
  \footnotesize
  \setlength{\tabcolsep}{3.5pt}
  \begin{tabular}{lrrrr}
    \toprule
    Concern category & $N$ & OpenCode & \makecell[r]{Claude\\Code} & Codex \\
    \midrule
    \multicolumn{5}{l}{\small\bfseries Presentation} \\
    \addlinespace[3pt]
    Structure / formatting    &  76 & \textbf{2} &  0 &  0 \\
    \midrule
    \multicolumn{5}{l}{\small\bfseries Correctness} \\
    \addlinespace[3pt]
    Executability             & 134 &  6 & \textbf{9} & \textbf{9} \\
    Consistency               & 111 & \textbf{8} &  2 &  0 \\
    Factual accuracy          & 199 &  8 & \textbf{10} &  6 \\
    Stale / outdated          &  69 & \textbf{5} &  4 &  4 \\
    Hallucinated / fabricated &  38 &  0 &  0 &  0 \\
    \midrule
    \multicolumn{5}{l}{\small\bfseries Coverage} \\
    \addlinespace[3pt]
    Conciseness               &  64 & \textbf{4} &  1 &  0 \\
    Specificity / actionability & 123 &  4 & \textbf{5} &  0 \\
    Completeness              & 648 & \textbf{27} & 22 &  5 \\
    \midrule
    \multicolumn{5}{l}{\small\bfseries Efficacy} \\
    \addlinespace[3pt]
    Discoverability / loading &  79 &  3 & \textbf{4} &  1 \\
    Wrong mechanism           &  20 & \textbf{1} &  0 &  0 \\
    Security                  &  38 & \textbf{1} & \textbf{1} &  0 \\
    \midrule
    \multicolumn{5}{l}{\small\bfseries Warrant} \\
    \addlinespace[3pt]
    Scope / placement         &  94 & \textbf{3} &  0 &  0 \\
    Audience fit              &  32 &  0 &  0 &  0 \\
    Triviality / low-value    & 146 &  0 & \textbf{1} &  0 \\
    Overreach / unwarranted   &  98 & \textbf{5} & \textbf{5} &  1 \\
    \midrule
    \textbf{Headline (unique concern)} & \textbf{1,912} & \textbf{75 (3.9\%)} & \textbf{62 (3.2\%)} & \textbf{25 (1.3\%)} \\
    \bottomrule
  \end{tabular}
\end{table}

Even a perfect three-way ensemble reaches only 6.9\% (132/1,912); the products miss largely \emph{different} concerns (triple-intersection $=$ 3 concerns), and 77.8\% of the 446 PRs (347) receive zero coverage from all three, so \textbf{the ceiling is a property of the \emph{task}, not of any one model.}

\subsection{Where the blindness falls}

The pattern in Table~\ref{tab:rq4-percat} is a floor drop as the concerns become more judgment-heavy. Reviewers reach 5--7\% on what current model knowledge can \emph{verify} (stale facts, executable errors, factual mistakes), but the judgment-heavy Warrant category at the bottom is \textbf{a near-total blind spot} (best 2.2\%, OpenCode; the others are below 2\%). The dividing axis is not hard-versus-easy; it is ``checkable against an external ground truth'' versus ``requires a normative judgment about whether the instruction should exist.'' This is the capability leg of RQ3's divergence, now confirmed under controlled effort: even a \emph{dedicated} reviewer, given the full repository, is blind precisely where human review is distinctive.

The few hits cluster in Executability, Factual accuracy, and Stale / outdated---AI succeeds mainly when review degenerates into ordinary code-or-fact checking, the competence cross-domain guardrails credit it with~\cite{naik2025crscore,pereira2026crbench} (see Discussion).

\emph{The misses are consequential, not overcaution.} Of the concerns all three reviewers missed, 71\% (1,173/1,653 with a recorded outcome) were ones the PR author subsequently acted on---rising to 77\% for the human-judgment categories---versus only 55\% of the concerns the AI \emph{did} catch. The AI does not preferentially recover the concerns that mattered most; if anything it misses them. (The gold's validation used author action as one well-foundedness signal, so the $\approx$70\% base rate is partly by construction; the load-bearing claim is the missed-71\% versus covered-55\% contrast.)

\subsection{The ranking is effort, not ceiling}

The recall ranking across products tracks \textbf{engagement, not capability}: Codex is last because 32\% of its reviews (141/446) are under 600 characters, single-sentence dismissals that never engage the instruction content. OpenCode and Claude write substantive, instruction-focused reviews and still land at 3.9\% versus 3.2\%.

\subsection{When the reviewer is confidently wrong}\label{subsec:rq4-cases}

Low recall would be benign if the reviewers stayed silent where unsure. They do not: a non-trivial share of the misses are not silence but active error, where the reviewer endorses the flaw, recommends the opposite of the human, or fabricates its own concern. Three hand-verified cases illustrate the modes.

\emph{False affirmation.} On \rr{Azure/azure-sdk-for-go#25394} the instruction file referenced a nonexistent legacy \texttt{services/} directory---the exact Hallucinated / fabricated concern the human flagged---yet OpenCode listed it as a positive finding (``Correct folder structure --- uses \texttt{sdk/} \ldots\ rather than the legacy \texttt{services/} directory''), praising the file on the very dimension it was wrong.

\emph{Contradiction.} On \rr{NVIDIA-NeMo/Curator#982} the author's explicit decisions were to focus on Python 3.12 and ignore GPU for now; Codex recommended the opposite on both, keeping 3.10 support and adding the GPU install index.

\emph{Fabrication.} On \rr{Azure/azure-sdk-for-java#44735} the human confirmed the file was correct (Java 21 is the current LTS), yet both Codex and Claude independently asserted ``Java 25 is the latest LTS'' and told the author to update---two model families converging on the same invented version.

These are illustrative, hand-verified cases, not a precise rate (Section~\ref{sec:method}). The direction matters: a reviewer that endorses flaws and fabricates facts does not merely fail to help---automated review of this kind would \emph{inflate} a file's apparent quality, a point we take up in the Discussion.

\begin{takeaway}
Three state-of-the-art agents recover only 1.3--3.9\% of validated human concerns even with full repository context, a task ceiling, not a model ceiling. The deeper problem is contamination: the reviewer loads the instruction file as its own context and, in several cases, obeys the very rules it is asked to audit.
\end{takeaway}

%% file: sections/09-discussion.tex
\section{Discussion}
\label{sec:discussion}

Our results describe a situation where most instruction files skip through their only quality gate twice. 
In most of the cases, instruction files don't receive reviews; even in cases they do, the reviews possibly leaving out the significant concerns. This double skip reflects a deeper mismatch between the emerging artifacts and the existing review process.

\subsection{The no-oracle gap as the structural blindness}

Instruction files lack an external oracle: unlike code, their correctness cannot be settled by compilers, linters, or test suites. 
This forces reviewers onto judgment-grounded terrain, where humans raise completeness, factual, and judgment-heavy concerns while AI gravitates toward surface structure (Figure~\ref{fig:asymmetry}). 
The result reveals a systematical blind spot: AI misses precisely the concerns that make human review indispensable (Section~\ref{sec:rq3}).
The blindness is structural, not informational. Today's agentic reviewers 

Worse, the file under review is self-executing: the design of the current agentic reviewers configures agents towards their review target (head ref of GitHub PRs), possibly introducing bias and undesirable behaviors. The to-be-reviewed agent instruction file acts as the agent's constitution, and a constitution cannot coherently question itself from an external standpoint. 
The convergence of independent products on the same blind categories, together with the contamination effects observed in our controlled trial, favors this structural reading over a pure knowledge-gap account (Section~\ref{sec:rq4}).

\subsection{AI reviewers on instruction files verses code}
\label{subsec:ai-reviewers}

Our results points out a major gap in the agents' reviewing capability in instruction files.
The three reviewer agents we tested, given full repository context, covered only \textbf{1.3\%}, \textbf{3.2\%}, and \textbf{3.9\%} of validated human concerns, in other words, missing out the majority.
The category-level pattern reinforces the interpretation: Warrant concerns were barely touched (best 2.2\%), hallucinated or fabricated content was detected 0\% of the time, and Triviality / low-value necessity concerns were caught only once out of 146 instances. Staleness emerged as the lone relative strength (6–7\%), suggesting that AI reviewers recover what can be checked against explicit, verifiable signals and miss what requires background knowledge typically preserved in developers' brains.

Yet, agents do know how to review code.
Comparable evaluations on reviewing source code report far higher recall: single tools in c-CRAB reach 20--32\% and a four-tool ensemble reaches 41.5\%~\cite{zhang2026ccrab}; Code Review Bench scores the strongest tool at 73--77\%~\cite{zverianskii2026codereviewbench}; even calibrated against CR-Bench, which sits near 5\% precision at high recall~\cite{pereira2026crbench}. Statistics from industrial tools agree—BitsAI-CR reports $\sim$75\% precision~\cite{sun2025bitsaicr}, AutoCommenter reports $\sim$40\% resolution ratio~\cite{vijayvergiya2024autocommenter}, and Copilot comments are 71\% actionable~\cite{github2026copilotreviews} (though reception is mixed~\cite{watanabe2024chatgptreview} and vendor benchmarks can be self-serving~\cite{deepsource2026benchmarks}). 

Taken together, the evidence indicates that, for agentic reviewers with production-level capabilities, reviewing instruction files remains a distinct and underserved challenge.
Whether the gap reflects a fundamental limitation or merely an under-represented target in current LLM training and evaluation is a question for future work; either answer would reshape how we design AI-assisted review.

\subsection{Towards instruction file review automation}
\label{subsec:tool-design}

The status of current reviews and the performance gap of current agents render reviewing instruction an emerging, unique, and significant workload. Derived from the insights of our study, we sketch three core characteristics for future designs on automating instruction file reviews.

\textbf{Tuned for instructions.} The reviewer should treat instruction files as executable policy, separate from documentation and configurations. That means recognizing defect classes specific to these files: hallucinated rules that do not match the repository, stale constraints inherited from earlier versions, trivial or low-value rules that add noise and cost, and unwarranted rules that over-reach the agent's scope. It also means resisting the structural bias toward surface well-formedness (completeness, structure, and style) that dominates current AI review (Section~\ref{sec:rq3}). Training or prompting should be grounded in the human-validated concern taxonomy in Section~\ref{sec:method}, with special attention to the judgment-heavy categories that human reviewers currently monopolize.

\textbf{Context-aware review.} Many of the human-only concerns cannot be judged from the diff alone. Whether a rule is stale, whether a convention is project-specific, and whether a restriction is warranted all require understanding the project's evolution and norms. The reviewer should proactively mine project context (pull-request history, issue discussions, comments, commit messages, and the current repository state) and derive project-specific quality checks from it. For example, a rule that mandates an outdated workflow step can be flagged by cross-referencing recent CI changes. A rule that duplicates an existing convention can be checked against prior agent-instruction PRs. This shifts the reviewer from a pattern matcher to a grounded investigator.

\textbf{Autonomous with escalation.} A useful reviewer should save developer effort without sacrificing quality. Surface defects such as formatting, broken internal links, duplicated rules, or simple factual mismatches can be fixed directly or explained in detailed, actionable comments. Hard-to-judge issues such as necessity, over-reach, hallucination, and project-specific convention should be escalated to a human, but not as a blank request. The agent should first mine the relevant context and then interview the maintainer in a structured way, to elicit the judgment needed for an informed decision rather than hallucinating a verdict. This division of labor matches what we observe: even on AI-authored instruction-file changes, 78\% still draw a human in for the final judgment calls (Section~\ref{sec:rq2}).

%% file: sections/08-threats.tex
\section{Threats to Validity}\label{sec:threats}

\textbf{Construct validity.} The 16-concern taxonomy was derived through inductive, identity-blind double-coding of 99.3\% of the 1,200 relevantly-reviewed PRs. The substantive tier-gate---whether a comment carries any concern---reached $\kappa = .918$ (Section~\ref{sec:method}); per-concern agreement is lower (macro-average $\kappa = .72$, most codes .71--.89, with a few judgment-heavy codes lower). The judgment-heavy concerns, i.e. Triviality, Overreach, Stale, Hallucinated, rest on few cases, so we anchor them to examples and interpret them cautiously (Section~\ref{sec:rq3}); the human/AI divergence, however, is driven by the high-$\kappa$, high-$N$ codes (e.g., Completeness, Factual accuracy), not these. Coverage and substance are reported on separate axes, so high AI coverage is never read as high substance (Section~\ref{sec:rq2}). Finally, the RQ4 benchmark measures \emph{recall} against a human gold: agents earn no credit for valid concerns raised outside it, so our claim is that they fail to \emph{recover human judgment}, not that they add no value; benchmark precision is left to future work.

\textbf{Internal validity.} Every reviewer-stratified result depends on labelling each account human, AI, or tool; the classifier validates against blind-coded gold at 95.3\% account-weighted and 98.5\% instance-weighted accuracy ($\kappa = .983$) (Section~\ref{sec:method}). Author-origin and the ride-along split are descriptive lower bounds: undisclosed AI use is uncounted, so the $\approx$1/3 AI-authored share is a floor, not a ceiling (Section~\ref{sec:method}). RQ3's observational gap could also reflect \emph{tasking} rather than \emph{capability}---wild AI accounts may never be asked to review substantively; RQ4 removes this confound by explicitly tasking agents with the full repository, and they still recover only 1.3--3.9\% (Section~\ref{sec:rq4}).

\textbf{External validity.} The corpus is gated to popular repositories (\texttt{star>100}, 2,246 repos) on a single English-dominant platform; the long tail and other forges are excluded. Since review is scarcer there, the 71.9\% no-substantive-review rate is a lower bound. Agent instruction files are a moving convention---our snapshot spans February~2021 to April~2026---so we frame the taxonomy and benchmark as exploratory rather than a stable base rate (Section~\ref{sec:method}; Section~\ref{sec:rq3}; Section~\ref{sec:rq4}).

\textbf{Reliability and reproducibility.} All coding disagreements were adjudicated to a single agreed gold label, so the labels we analyze are reconciled, not disputed. The gold set holds 1,962 validated concerns across 468 PRs (1.26\% false positive); the RQ4 benchmark scores the 1,912 concerns across the 446 PRs reconstructable at review time. Coverage is scored by two human judges, and the product ranking is stable under a stricter independent re-judging (Section~\ref{sec:rq4}); we treat the point estimates as indicative, not precise, given a \emph{single, nondeterministic} run per product. Two residual threats cut in our favor: both identity-blind \emph{leakage} and \emph{training-cutoff contamination} could only \emph{inflate} recovery, so our figures upper-bound true capability---and RQ4's fabricated-fact cases (e.g., an invented ``Java~25 LTS'') show agents guessing, not recalling. 

%% file: sections/10-conclusion.tex
\section{Conclusion}\label{sec:conclusion}

Agent instruction files are executable policy with no per-change oracle: they slip past compilers, type-checkers, and test suites. Review is their only quality gate. Across 2,870 pull requests from 2,246 repositories, we found that gate is largely broken, with over 70\% of instruction-file changes receiving no substantive review. When AI reviewers fill the gap, they are structurally blind to the concerns that most need human judgment: they catch surface structure but miss factual errors, necessity, staleness, and hallucinated content. This is not a generic ``AI cannot review'' finding; the same agents review source code competently, but their blindness here is specific to executable policy.

We contribute the first taxonomy of instruction-file review concerns (16 categories), which exposes the human/AI divergence, and a benchmark of 1,962 human-validated concerns across 468 PRs against which three widely-used agents recover only 1.3--3.9\%. The no-oracle gap drives this divergence: without ground truth for correctness, human and AI reviewers attend to non-overlapping concerns, and because an agent-native reviewer loads the file as its own context, the artifact under review can shape the very agent that audits it. A constitution cannot coherently question itself. An instruction-file change is a change to agent behavior, not a documentation edit. Whether the gap reflects a fundamental limitation or an under-represented training target remains open; either answer would reshape how we design review for executable policy.